# REAL-TIME STRUCTURAL HEALTH MONITORING SYSTEM USING INTERNET OF THINGS AND CLOUD COMPUTING

H. Chang[1] and T. Lin[2]

## ABSTRACT

Real-time monitoring on various structural behaviors, particularly displacement and acceleration, serves important and valuable information for people; for example, they can be used for active control or damage warning. With recent advancement of Internet of Things (IoT) and client-side web technologies, wireless integrated sensor devices nowadays can process real-time raw sensor signal data into target measurements, such as displacement, and then send the results through a standard protocol to the servers on the Internet (i.e., the cloud). The monitoring results are further processed for visualization purpose in the servers and the computed results are pushed to connected clients like browsers or mobile applications in real-time. We build a real-time cloud-based system that can receive heterogeneous IoT data, allow users to create a three dimensional (3D) model online according to the real world structure, and the monitoring results can be visualized in that model. In this paper, we illustrate the software architecture of the proposed system and focus on the technologies that are used, like client-side scripting, NoSql database, and socket communication. We also present the challenges of displaying the overall movement and shape transformation of the 3D structural model. Thus, each internal connected element's rotations and translations are obtained by converting the monitoring results of each sensor device measured in global coordinate system. To overcome this, we create the inverted movement calculation method. A simple 3D two-level structural model and simulated sensor displacements are use to demonstrate system function and validate the inverted movement calculation method.

[1]Research Scientist, Link Dynamic, Irvine, CA 92604 (email: hungfuc@gmail.com)
[2]Professor, Dept. of Civil Engineering, National Chiao Tung University, Taiwan (tklin@nctu.edu.tw)

# Real-time Structural Health Monitoring System Using Internet of Things and Cloud Computing

H. Chang[1] and T. Lin[2]

## ABSTRACT

Real-time monitoring on various structural behaviors, particularly displacement and acceleration, serves important and valuable information for people; for example, they can be used for active control or damage warning. With recent advancement of Internet of Things (IoT) and client-side web technologies, wireless integrated sensor devices nowadays can process real-time raw sensor signal data into target measurements, such as displacement, and then send the results through a standard protocol to the servers on the Internet (i.e., the cloud). The monitoring results are further processed for visualization purpose in the servers and the computed results are pushed to connected clients like browsers or mobile applications in real-time. We build a real-time cloud-based system that can receive heterogeneous IoT data, allow users to create a three dimensional (3D) model online according to the real world structure, and the monitoring results can be visualized in that model. In this paper, we illustrate the software architecture of the proposed system and focus on the technologies that are used, like client-side scripting, NoSql database, and socket communication. We also present the challenges of displaying the overall movement and shape transformation of the 3D structural model. Thus, each internal connected element's rotations and translations are obtained by converting the monitoring results of each sensor device measured in global coordinate system. To overcome this, we create the inverted movement calculation method. A simple 3D two-level structural model and simulated sensor displacements are use to demonstrate system function and validate the inverted movement calculation method.

## Introduction

Civil structures in general request high investment and are expected to have a long service life [1]. Each structure is often unique regarding its material and shape, and its behaviors often change due to their age, usage or environmental factors. Key structures, such as bridges, power utilities, nuclear power plants, and dams, particularly requires continuously monitor structural current condition in order to support management decisions and provide necessary maintenance in time. Structural Health Monitoring (SHM) aims to develop a system that can continuous monitor a structure, delivery current structural responses, and even alert when the structure passes the design domain. A more effective SHM should provide monitoring results in real-time and online; therefore, immediate response can be carried out to avoid further loss and damage of the structure and detailed structural behavior data can be used in design or study in the future.

Sensors, a data processing component and a health diagnosis unit are three major parts in a typical SHM system [2]. With recent advance in technology, industrial developments and research in these three domains have been shifted to apply latest standards and software libraries. In the area of sensor or sensor network, traditional wireless sensor network has been moved to Internet of Things (IoT) because wireless sensor network is developed in a close system that needs to use gateways to connect or to export sensor data to the external world or the Internet [3]. Things in IoT, often called smart objects or devices, use standard communication protocols, like Hypertext Transfer Protocol (HTTP) or Message Queue Telemetry Transport (MQTT) to interact among each other and to communicate server systems across platforms over Internet [4] [5]. Regarding data process and health evaluation, after cloud computing technology emerges, systems can be implemented and distributed across various regions, messages can be transmitted over the HTTP by web services, and data can be stored, processed, and accessed via websites or mobiles.

[1]Research Scientist, Link Dynamic, Irvine, CA 12345 (email: hungfuc@gmail.com)
[2]Professor, Dept. of Civil Engineering, National Chiao Tung University, Taiwan (tklin@nctu.edu.tw)

In this project, our approach is developed based on IoT sensor devices and cloud computing in real-time fashion. Our SHM system visualizes real-time dynamic structural behaviors in 3D model through web or mobile so users can easily understand and observe the entire structural movements. Since sensors at their installation locations measure the structural three-dimensional displacements, how to show entire structural transformation on its 3D model at every display time interval depends on converting the local displacement data from its elements into the overall movement in global coordinate system. Hence, we create the inverted movement calculation method to output the 3D structure's movements in global coordinate system from sensor displacement data inputs.

## Real-time Structural Health Monitoring System

### System Architecture

There are several aspects in our real-time SHM system design. First, a web-based user interface is built across platforms and targets for two primary user groups, project administration and monitoring users. Project administration users manage sensor device details (e.g., locations and identification number) and build 3D models according to the real structure. Monitoring users observe measurements that reflect on the 3D model. Second, all real-time monitoring data are collected from IoT sensor devices via wireless network. Third, health evaluation is calculated in the servers on the Internet based on the 3D structural model, monitoring data, and sensor deployment location. Last, a user client also performs real-time communications with servers to show 3D model changes at the client side.

There are two kinds of servers in our real-time SHM system (see Fig. 1). One is the regular HTTP server that does all the computations, interacts between user's browsers and itself, and saves monitoring results, 3D model, and sensor information into a database. IoT sensor device's monitoring results are pushed to this server according to its sampling frequency via HTTP protocol. The other kind serves as a real-time message server that is implemented by JavaScript Socket.io library and Node.js framework. This message server pushes the computed monitoring results received from HTTP server to user's browser. These monitoring data can be used to move and rotate each 3D element inside the structural model in real time; as a result, the whole 3D structural model dynamic behaviors can be displayed on the browser.

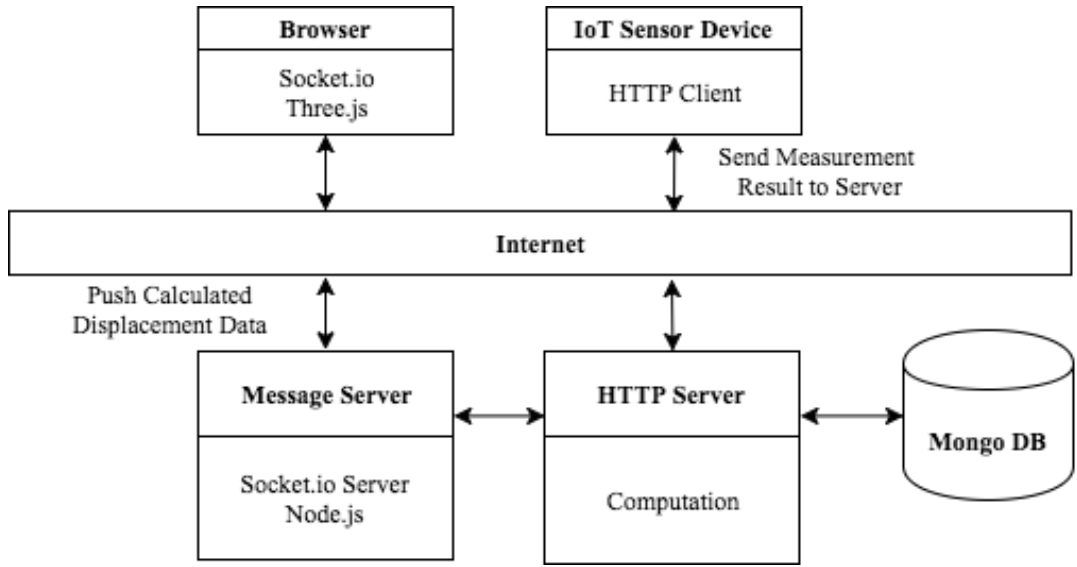

Figure 1.  Software architecture of real-time SHM system

**Significant components of implementation**

Web technology has been widely used in modern software systems to facilitate inter-operability between server-side and client-side program or to exchange data among servers, and there are many open source software libraries available. Because our objective is to host scalable services in the cloud to be accessed by different kinds of devices or software platforms, multiple related open source web-based libraries are applied. In terms of the real-time communication, we use Socket.io that has a client-side component running in the browser and a server-side component on top of Node.js to send bi-directional messages between client and server based on pre-defined events. The system saves data in MongoDB, which is classified as a document-oriented, so called NoSQL, database. To render 3D building model's changes on the client-side, Three.js is employed. These open source libraries build the key components of the proposed system.

**Sensor Device**

The IoT sensor device that we created for this proposed approach contains a microcontroller chip, ESP8266, and an ADXL345 accelerometer. ESP8266 is a commercial IoT microcontroller that provides wireless capability and small size flash memory. ESP8266 also has station infrastructure (STA) mode that sets a service set identifier (SSID) to allow end user computers to connect to its wireless network. When a user's browser links to the SSID's network and obtains a local IP address during STA mode, a web user interface for configuring the device is shown for users. Inside this configuration page, there is a unique identification number. Therefore, the HTTP server can use this number to identify the device, associate with its location, and then do computation for 3D model visualization.

The IoT device itself also does simple computation to convert the measured raw signal into meaningful data like acceleration 0.5g at x-axis. It buffers the data and then pushes the data to our HTTP server so that the multiple monitoring records that associate with the device's identification number can be saved into database in one request.

## 3D Structural Model Visualization Computation Method

**Movement and Rotation**

Transformations and rotations of a geometry model involve its element's local and global coordinate systems. The origin of local coordinate system is the center of the object. The object's transformation is calculated by moving the center of the object along with its local x-, y-, or z-axis and its rotation is based on the rotation of local coordinate system. In Fig. 2, axis $X_G$ and $Y_G$ belongs to global coordinate. Before the 3D object moves, its local origin is the same as the global origin (i.e., $O_L = O_G$) and the local coordinate system is consistent with the global coordinate system (i.e., $X_L$ is the same direction as $X_G$ and $Y_L$ is the same direction as $Y_G$). When a translation occurs, the local origin move to new location. The local origin moves from (0, 0) to a new point ($D_X$, $D_Y$) at the global coordination.

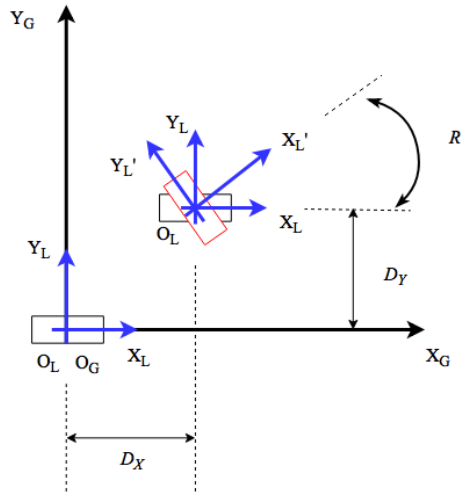

Figure 2. Two dimensional movement caused by X-Y forces

A rotation has similar effects as translation. But, a rotation is always at the center point of the object that refers to its local coordinate. In Fig. 2, the object rotates R degrees at the center according to local Z-axis so the local X and Y axis also changes accordingly and make $X_L$ ($Y_L$) have different direction from $X_G$ ($Y_G$).

Therefore, to render a movement, each element must rotate and translate from its previous position in global coordinate system because change of the model cannot be simply re-drawn by connecting the node to node according to each node's global coordinates. At each measurement moment, the system can only obtain each sensor node's position in global coordinate system by composing the displacement at each axis. The system needs to convert the three dimensional global displacements into a rotation and a translation in local coordinate system each time.

**Inverted Movement Calculation Method**

The proposed inverted movement calculation method takes three dimensional displacements that are measured every timestamp at each sensor node inside real world structure and outputs the rotation and translation values in local ordinate system for the structure's 3D model. The inverted movement calculation method now only focuses on column's movements. Therefore, there are two assumptions in this method. First, all the columns of the structure are rigid bodies; which means, there is no deformation during rotation and translation. Second, floor or beam can be extended along with the moving direction and there is no lateral deformation or bending.

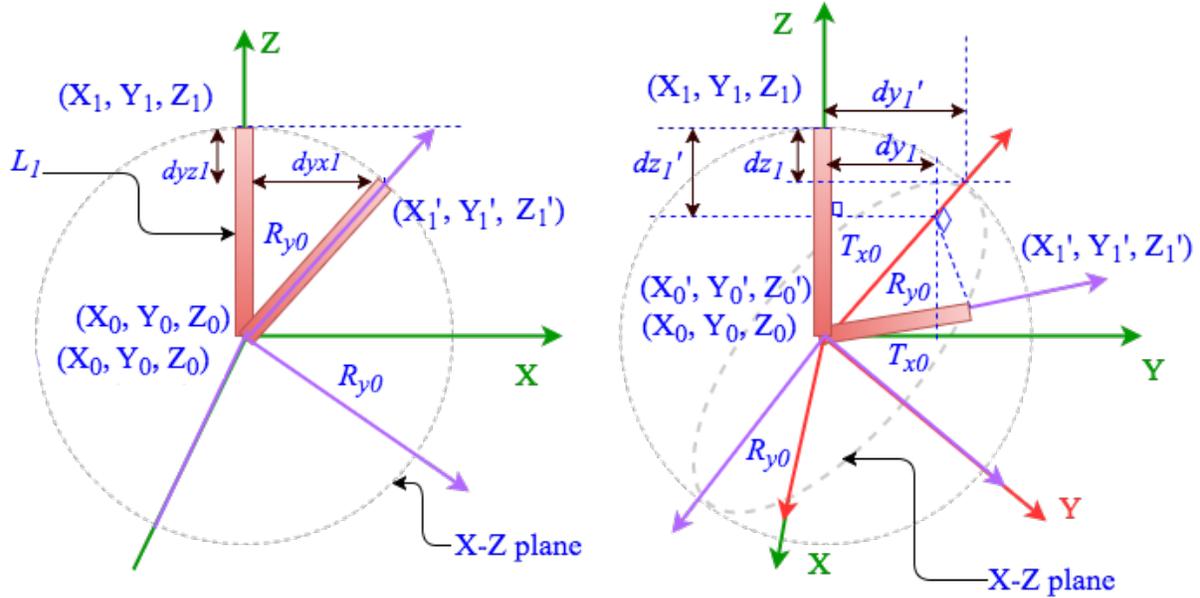

Figure 3. Two dimensional rotation caused by X and Y forces

Given that columns are connected to each other from column 1 to n, Fig. 3 shows how we derive the formula for the first column (i.e., column 1) in the method. The column 1, whose length is $L_1$, rotates $T_{x0}$ degree around x-axis, and then $R_{yo}$ around y-axis from initial vertical position. Eq. 1 shows the way to get rotation angle $R_{y0}$. $(X_0', Y_0', Z_0')$ and $(X_1', Y_1', Z_1')$ are bottom and top points of the column, respectively, after two rotations. Since the rotation around x-axis does not make any displacement at x-axis, x-axis displacement only occurs while the column rotates around y-axis. As a result, $d_{yx1}$ equals to $(X_1' - X_0')$.

$$R_{y0} = \sin^{-1}(d_{yx1} / L_1) = \sin^{-1}((X_1' - X_0') / L_1) \tag{1}$$

In Eq. 2, the final displacement at y-axis $d_{y1}$ equals to $(Y_1' - Y_0')$ and also equals to $L_1 \cos(R_{y0}) \sin(T_{x0})$. So, $T_{x0}$ can be calculated by using the final y-axis displacement and the x-z plane rotation angle $R_{y0}$. When $L_1$ is projected firstly by $R_{y0}$ and then by $T_{x0}$ degrees, $Z_1'$ can be derived (see Eq. 3). The final z-axis displacement $d_{z1}'$ that is contributed by both rotations can be gotten from Eq. 4.

$$T_{x0} = \sin^{-1}((d_{y1}) / (L_1 \cos(R_{y0}))) = \sin^{-1}((Y_1' - Y_0') / (L_1 \cos(R_{y0}))) \tag{2}$$

$$Z_1' = L_1 \cos(R_{y0}) \cos(T_{x0}) \tag{3}$$

$$d_{z1}' = L_1 (1 - \cos(R_{y0}) \cos(T_{x0})) \tag{4}$$

Eq. 1 to 4 can be generalized as Eq. 5 to 8. In these equations, $R_{yn-1}$ is the rotation angle around y-axis of column n, $T_{xn-1}$ is the rotation angle around x-axis of column n, $L_n$ the length of the column n, $(X_{n-1}', Y_{n-1}', Z_{n-1}')$ and $(X_n', Y_n', Z_n')$ are coordinates of bottom and top points of the element n, and $d_{zn}'$ is the z-axis displacement of column n. If we have multiple connected columns and two rotations at each column, we can get two rotation angles and z-axis displacement from Eq. 5 to Eq. 8.

$$R_{yn-1} = \sin^{-1}((X_n' - X_{n-1}') / L_n) \tag{5}$$

$$T_{xn-1} = \sin^{-1}((Y_n' - Y_{n-1}') / (L_n \cos(R_{yn-1}))) \tag{6}$$

$$Z_n' = L_n \cos(R_{yn-1}) \cos(T_{xn-1}) \tag{7}$$

$$d_{zn}' = L_n (1 - \cos(R_{yn-1}) \cos(T_{xn-1})) \tag{8}$$

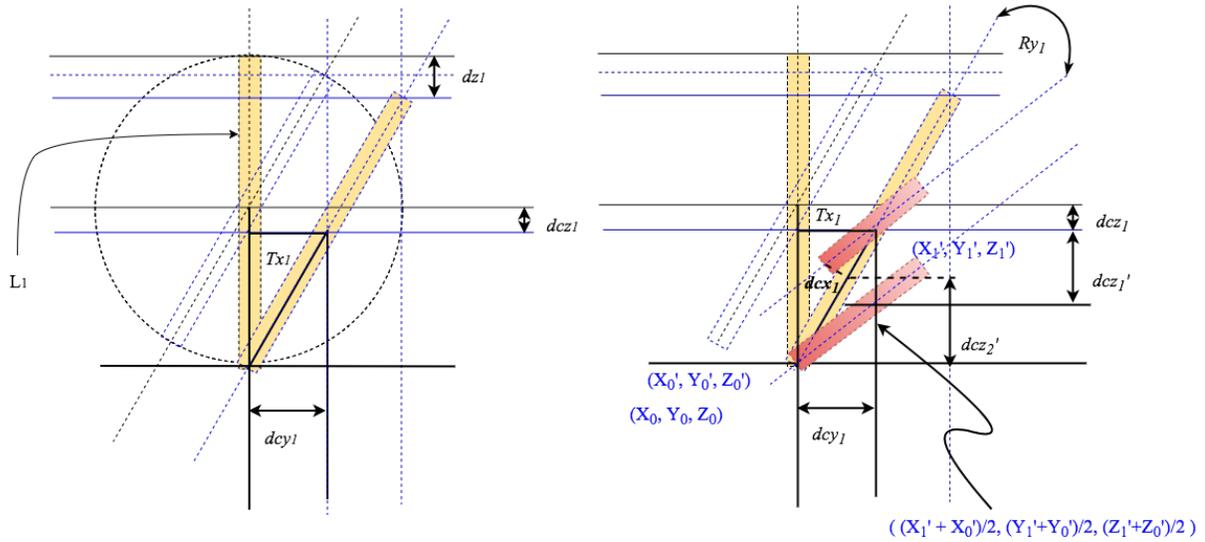

Figure 4.  Two rotations at center point of the column

Unlike rotation at the bottom, 3D model rotation in the Three.js actually happens at the center of the object. Therefore, to make the same rotation and movement at the bottom like Fig. 3, the column object can firstly rotate same angle at the center and then move the bottom point to the original bottom position (see Fig. 4). For example, making x-axis rotation at center includes the y-axis displacement $d_{cy1}$ that is caused by $T_{x1}$ degree rotation and the bottom point displacement from point $Y_1'$ to $Y_1$. In details, for element n rotating around x- and y-axis, total x-axis displacement at the center point $d_{cxn-total}$ equals to sum of displacement $d_{cxn}$ and bottom point displacement $(X_{n-1}' - X_{n-1})$. Total y-axis displacement $d_{cyn-total}$ equals to sum of displacement $d_{cyn}$ and bottom point displacement $(Y_{n-1}' - Y_{n-1})$. Total z-axis displacement $d_{czn-total}$ needs to consider the movements that are made by x and y rotation $d_{czn} + d_{czn}'$ and bottom point movement $(Z_{n-1}' -$

$Z_{n-1}$). As a result, total displacement equations are shown as following.

$$d_{cxn\text{-}total} = (1/2)\, L_n \sin(R_{yn}) + (X_{n-1}' - X_{n-1}) \tag{9}$$

$$d_{czn\text{-}total} = d_{czn} + d_{czn}' = (L_n/2)\,((1-\cos(T_{xn})) + (1-\cos(R_{yn}))\cos(T_{xn})) + (Z_{n-1}' - Z_{n-1}) \tag{10}$$

$$d_{cyn\text{-}total} = (L_n/2)*\sin(T_{x1}) + (Y_{n-1}' - Y_{n-1}) \tag{11}$$

**Sensor Deployment**

Each IoT sensor device measures displacements at installation point of the structure. It is also the node of the rotation and displacement in the 3D structural model. Our inverted movement calculation method has a rigid body assumption. To express the real world column bending, equipping more sensor devices on the column is highly suggested. In Fig. 5, we can discover the difference shown on the 3D structural model between two kinds of sensor deployments. Installing more sensor devices can present more accurate movement in the 3D model. However, at least, installing sensor device at the joints in the real world structure is necessary.

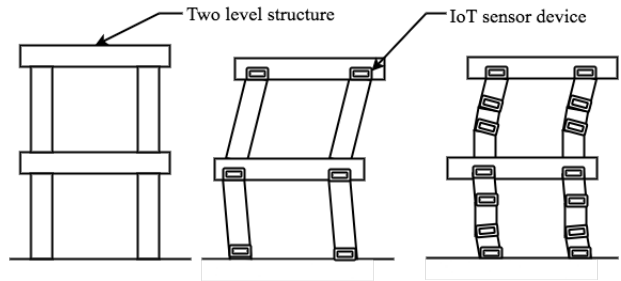

Figure 5. Deformation difference due to different sensor deployment

**Implementation and Validation**

**User Operations**

Management and monitoring are two major user operations in our real-time SHM system. Before users use the system, they install sensor devices at the real world structure that is going to be monitored. Then, they edit the 3D structural model according to that structure, and save the sensor information (e.g., location) into the system. Sensor location is critical to the 3D model movement calculation; therefore, the x, y, z coordinates of each sensor must be measured according to the real length, width and height of each elements of the structure and be properly scaled in the 3D model if necessary. When there is a new sensor device that is installed in the same structure, users just need to update the sensor information in the system. Like plug and play, the monitoring results will be automatically calculated based on updated sensor information.

The implementation allows multiple structures to be monitored. Hence, users firstly need to select the structure that they want to observe and then they can view the real-time movement of that structure. The maximum displacement at each axis is also set when users edit the structural model and sensor information. Warning message will be sent if any movement of the structure exceeds the maximum displacement. To note that, updating any sensor information will immediately reflect on the movement calculation in real-time without user's awareness.

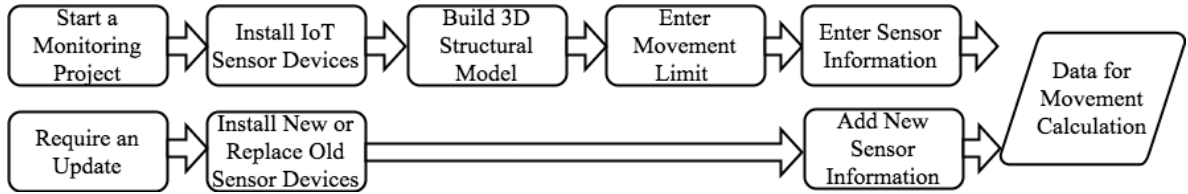

Figure 6. Management operations

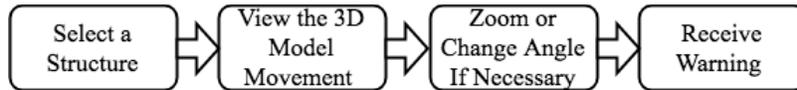

Figure 7. Monitoring operations

**Sensor Device**

Like regular IoT devices, the sensor device can connect to the Internet via a typical wireless router and becomes an independent computing resource to exchange messages with servers. Users can use browser to configure the sensor device. Having device IP address indicates that the sensor device already connects to the Internet (see Fig. 8). The identification number (i.e., Device ID) on this page should be input into our SHM system so the server can recognize which monitoring results are from this device.

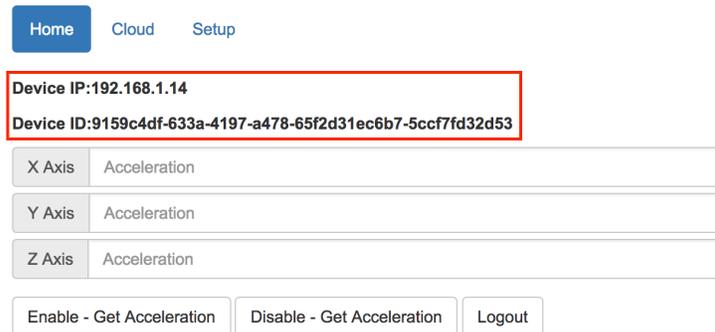

Figure 8. User Interface of Sensor Device

**Validation**

Two servers of the real time SHM system are implemented and hosted in the one core CPU, 512MB ram, and Ubuntu OS machine on the Internet. To validate our proposed inverted movement calculation method, we build a simple two-layer structure and assign one sensor at each joint. Therefore, there are totally eight devices that are input into the system. We perform a simulation by delivering the mock displacements from each sensor at the sampling rate 20 Hz to the server.

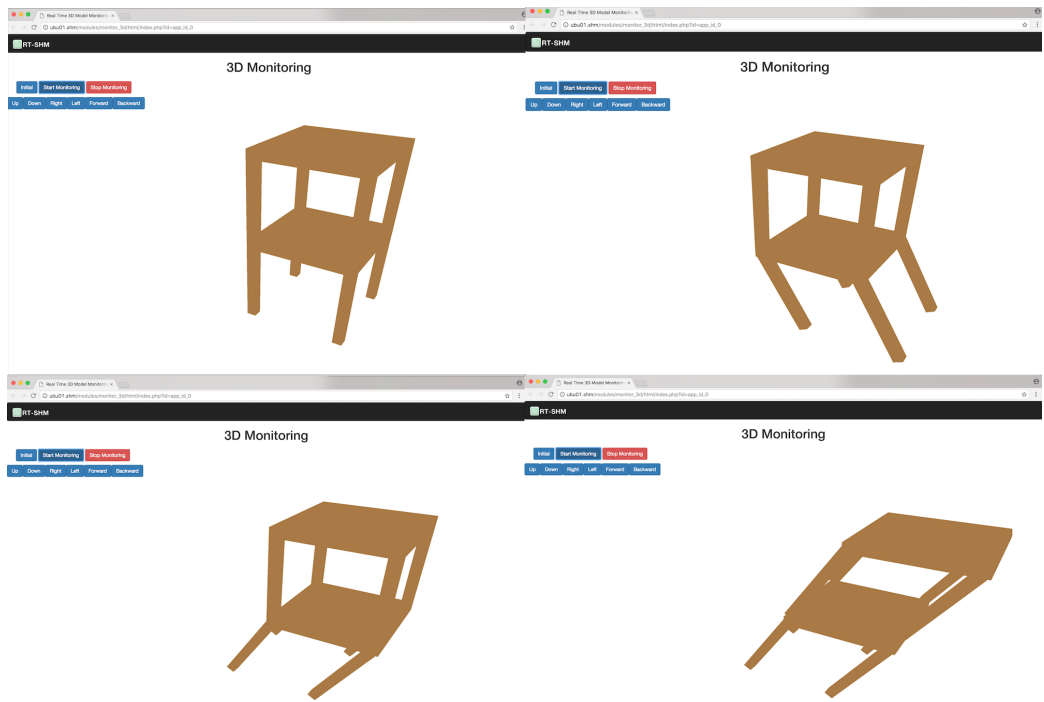

Figure 9. Real time SHM system demonstration

Fig. 9 shows the some movements of the two-level structural model in a browser. During the monitoring, the system displayed a movement every five seconds. Neither the data process in the server nor communication between servers and IoT sensor devices cause any delay. The system provides function to change the view angle and to zoom in or out. Hence, we decide to simply use observation from various angles to validate the purposed method. Throughout the whole simulation, there is no disjoint between elements or wrong rotations; that is, the system indeed displays movements correctly. As a result, the inverted movement calculation method should approximately reproduce the actual movement and deformation of real world structure.

## Future Work

As a next step we will perform an experiment on a real world structure with more IoT sensor devices; therefore, we will be able to compare the real movement and shape transformation with its 3D model. We also want to improve the inverted movement calculation method by dropping the rigid body assumption and considering a curve between two different nodes. Furthermore, the local damage detection can be introduced into this system so it will not be only limited to damage warning about exceeding maximum structural displacement.

## Conclusions

Typical SHM systems provide monitoring result that can alert when the structure passes the design capacity or even can detect possible damage. We proposed a real time SHM system that is based on IoT and web technologies to provide immediate responses to people through the Internet. The system lets users build the 3D structural model online and display its dynamic movements. We also create the inverted movement calculation method that converts sensor device's local three-dimensional displacements into overall 3D structural model's global movement. From the demonstration, the proposed system and method can show correct movement without any delay.